\documentclass[twocolumn,showpacs,preprintnumbers,amsmath,amssymb,prb,aps]{revtex4-1}
\usepackage{epsfig}
\usepackage{epstopdf}
\usepackage{graphicx}
\usepackage{dcolumn}
\usepackage{bm}
\usepackage{slashbox}
\usepackage{textcomp}
\usepackage{array}
\usepackage{subcaption}
\usepackage{booktabs}

\begin{document}

\title{Instrument for simultaneous measurement of Seebeck coefficient and thermal conductivity in the temperature range $300-800$ K with python interfacing}
\author{Shamim Sk$^{1,}$}
\altaffiliation{Electronic mail: shamimsk20@gmail.com}
\author{Abhishek Pandey$^{1}$}
\author{Sudhir K. Pandey$^{2,}$}
\altaffiliation{Electronic mail: sudhir@iitmandi.ac.in}
\affiliation{$^{1}$School of Basic Sciences, Indian Institute of Technology Mandi, Kamand - 175075, India}
\affiliation{$^{2}$School of Engineering, Indian Institute of Technology Mandi, Kamand - 175075, India}


\begin{abstract}
Fabrication and characterization of instrument for  high-temperature simultaneous measurement of Seebeck coefficient (S) and thermal conductivity ($\kappa$) has been carried out with  python automation. The steady-state based Fourier's law of thermal conduction is employed for $\kappa$ measurement. The parallel thermal conductance technique is implemented for heat loss measurement. Introducing the thin heater and insulating heater base minimize the heat loss and make the way easier to arrive at high temperature. Measurement of S is carried out using differential method. Same thermocouples are used to measure temperature as well as voltage for S measurement. Care of temperature dependent S of thermocouple has also been taken. Simple design, small size, lightweight make this instrument more robust. All the components for making sample holder are easily available in the market and can be replaced as per the user demand. This instrument can measure samples with various dimensions and shapes in the temperature range $300-800$ K. The instrument is validated using different class of samples, such as nickel, gadolinium, Fe$_{2}$VAl and LaCoO$_{3}$. Wide range of S from $\sim-$20 to $\sim$600 $\mu$V/K and $\kappa$ from $\sim$1.1 to $\sim$23.5 W/m-K are studied. The measured values of S and $\kappa$ are in good agreement with the reported data.     


\end{abstract}

\maketitle
\section{INTRODUCTION}
Fast consumption of natural energy sources (e.g. coal, oil, gas etc) has always pushed the researchers to find the alternative sources of energy. In this contest, thermoelectric generator (TEG) is one which can convert the industrial waste heat into useful electricity.\cite{teg1, teg2} TEG is characterized by the means of dimension-less parameter, called \textit{figure-of-merit} (ZT),\cite{zt}
\begin{equation}
ZT = \dfrac{S^{2}\sigma T}{\kappa},
\end{equation}
where, \textit{S}, $\sigma$ and $\kappa$ are the Seebeck coefficient, electrical conductivity and thermal conductivity, respectively. \textit{T} is absolute temperature of the sample. The efficient TE materials should possess high S and $\sigma$ with low $\kappa$ to attain a high ZT. To calculate ZT, these three parameters have to be measured. Out of these, $\kappa$ is most difficult quantity to measure specially at high temperature region due to undefined amount of heat loss by conduction, convection, and radiation. From Eqn. 1, it is seen that only the quantity S enters in ZT as square. Therefore, any small error in measuring S can give rise to the significant change in ZT value. Hence, precise measurement of S is required to get more accurate ZT.   

The S is simply defined as the production of electrical voltage in the presence of temperature gradient. S can be measured in two ways: integral and differential. In integral method, the temperature of one end of the sample is varied, whereas other end of the sample is kept at constant temperature. The extra cooling system is required to maintain one end of the sample at constant temperature. This makes the instrument more complex and costly. This method is also not suitable for nondegenerate semiconductors and insulators.\cite{integral1, integral2} In the differential method, S of the sample is calculated by: $S = -\frac{\Delta V}{\Delta T} + S_{wire}$. Where, $\Delta V$ is the induced voltage difference across the temperature difference $\Delta T$ within a sample, and $S_{wire}$ is the Seebeck coefficient of connecting wire. Most of the seebeck measuring instruments prefer to use this method because of its simplicity and no additional requirement of cooling system.\cite{differential1, differential2, differential3, differential4} In this method, the connecting wire and thermocouple should be attached at the identical point of the sample to measure temperature and induced voltage, which is the main difficulty of this method.\cite{integral2} To overcome this issue, Boor \textit{et al.}\cite{boor} came with the different appraoch under differential method, where same thermocouple is used for the measurement of temperature as well as induced voltage. Later, this method of Boor \textit{et al.} is used by others\cite{patel_seebeck,saurabh_seebeck} to measure S from 300 K to 620 K. Further, Kumar \textit{et al.}\cite{saurabh_seebeck2} have utilized this method for measuring S in the temperature range $80-650$ K. The difficulty in implementation of this method for high temperature application comes from inserting the thermocouple for measuring output signal. Here, we have fabricated the instrument using the method proposed by Boor \textit{et al.}\cite{boor} to measure S in the temperature region $300-800$ K. We have fixed the thermocouple with the help of fire gun and using thin wire of the special designed material contains 95$\%$ silver and 5$\%$ zinc. Details of welding of thermocouple is described in Sec. III. 

In most general, the $\kappa$ is defined as the ability of materials to conduct heat. Various methods have been developed for the measurement of $\kappa$ with their own merit and demerit.\cite{tong} The accurate measurement of $\kappa$ is always a terrible job. To quantify the amount of heat flowing through the sample is really difficult. The existing $\kappa$ measurement techniques are broadly classified into two divisions: steady-state and non-steady-state. In steady-state, measurement is carried out when the temperature of the material at a paticular position does not vary with time. The non-steady-state based method relatively takes less time for the measurement as this method does not need to wait for steady-state. The laser flash method developed by Parker \textit{et al.}\cite{parker} is most popular among the non-steady-state based techniques. But, this method has been tested for metals only around room temperature and proposed that measurement can be done for all types of solid materials in any temperature range. Another popular technique is 3$\omega$ method, which has been developed to measure the thermal conductivity of thin film and thin wire mainly.\cite{tong,bourlon,lu} In this method, a frequency of $\omega$ is used to excite heater and then $\kappa$ can be measured from the 3$\omega$ response. 

The signal analysis in non-steady-state method is quite heavy, while steady-state method involves easy signal analysis. The available steady-state based methods are axial heat flow, guarded hot plate, direct electrical heating, hot wire method etc.\cite{tong} In steady-state method, one-dimensional Fourier's law of heat conduction is employed to measure $\kappa$. The measurement accuracy of this method totaly depends on how accurately we measure the amount of heat passing through the sample. Because, heat is lost through conduction, convection, and radiation, which becomes more significant at high temperature.\cite{amatya} The undefined amount of heat loss is measured by different ways. Zawilski \textit{et al.}\cite{zawilski} described a parallel thermal conductance technique for measuring $\kappa$. In this technique, heat loss is determined by running the instrument without a sample. They measured $\kappa$ from room temperature to 12 K. Later, Dasgupta \textit{et al.}\cite{differential2} implemented this technique to measure $\kappa$ from room temperature to 700 K. Wide variation of $\kappa$ is reported at room temperature only. They reported the $\kappa$ value of CrSi$_{2}$ from room temperature to 650 K with the maximum value of $\sim$2.2 W/m-K. Recently, Patel \textit{et al.}\cite{patel_kappa} fabricated the instrument for measuring $\kappa$ using the same technique of parallel thermal conductance. But their instrument is limited to measure $\kappa$ from room temperature to 620 K. Needless to say that heat loss increases with increase in temperature and measurement of $\kappa$ at high temperature becomes more challenging. Keeping this in mind, we have fabricated the instrument by making very thin heater and insulating heater base to minimize the heat loss. Instrument is tested from 300 to 800 K for measuring wide range of $\kappa$ with standard samples. The speciality of this instrument is that it can measure S as well as $\kappa$ simultaneously.  

Manual data collection is always a painful job for a researcher. This is more troublesome when measurement is carried out for a longer time. In this regard, we can say that manual monitoring of S and $\kappa$ is a hectic and time taking process when these measurements are performed in steady-state and wide temperature range. In the present study, our instrument facilitates the simultaneous measurement of S and $\kappa$ in steady-state and wide temperature range. Here simultaneous measurement saves time, but it requires the number of output signals collection by maintaining the thermal stability, which increases the difficulty for an user. Therefore, to make the job easier, automation of measurement can be done where these complexities come into account. One of the popular interfacing tool is LabVIEW. But, LabVIEW is a commercially paid graphical language. So, to avoid the commercial cost, we have automated the measurement with the open source "python" programming language. Python has vast libraries support which evades the dependency of external libraries.    

In this work, we have designed and fabricated the simple and low cost instrument for the simultaneous measurement of S and $\kappa$ in the temperature range $300-800$ K. Differential method has been implemented for measuring S, whereas parallel thermal conductance technique is used for the measurement of $\kappa$. Every parts of the sample holder are easily available in the market and can be replaced as per the user interest. This instrument can measure samples with various dimensions and shapes. Instrument is tested by performing the measurement on the standard samples of nickel, gadolinium, Fe$_{2}$VAl and LaCoO$_{3}$. A wide range of S from $\sim-$20 to $\sim600$ $\mu$V/K and $\kappa$ from $\sim$1.1 to $\sim$23.5 W/m-K are measured using our instrument. The measured values give good match with reported data. The instrument is automated with open source programming language python to minimize the human effort.

\section{FORMULATIONS OF MEASUREMENT}
In this section, we discuss the formulations used for the measurement of S and $\kappa$ in the present study. 

\subsection{Measurement of Seebeck coefficient}
Here, the differential method proposed by the Boor \textit{et al.}\cite{boor} has been implemented to measure S. In this method, S of the sample is calculated by 
\begin{equation}
S = -\frac{U_{neg}}{U_{pos} - U_{neg}}S_{TC}(\overline{T}) + S_{neg}(\overline{T}),
\end{equation}  
where $U_{pos}$ and $U_{neg}$ are the voltages measured across the positive legs and negative legs of thermocouple wires, respectively. $S_{TC} = S_{pos} - S_{neg}$, is the Seebeck coefficient of thermocouple, where $S_{pos}$ and $S_{neg}$ are the Seebeck coefficient of positive legs and negative legs of the thermocouple wires, respectively. $\overline{T} = (T_{1} + T_{2}$)/2 is the absolute temperature, where $T_{1}$ and $T_{2}$ are hot side and cold side temperature. Two K-type thermocouples are used to measure $T_{1}$, $T_{2}$, $U_{pos}$ and $U_{neg}$ as shown in Fig. 1. Here, it is important to note that temperature dependency of $S_{TC}$ and $S_{neg}$ have been taken care in the present study.

\begin{figure}
\includegraphics[width=0.85\linewidth, height=5.5cm]{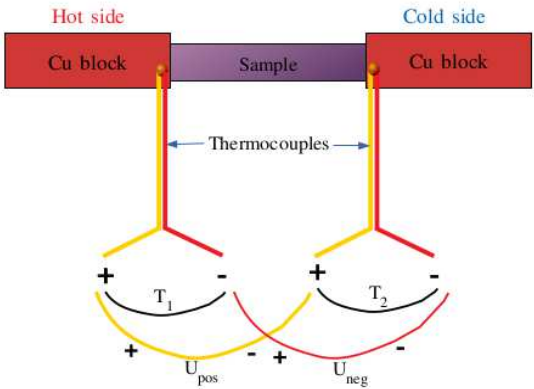} 
\caption{\small{Wiring scheme for the measurement of $T_{1}$, $T_{2}$, $U_{pos}$ and $U_{neg}$ using thermocouples.}}
\end{figure}

\begin{figure*}
\includegraphics[width=0.9\linewidth, height=7.0cm]{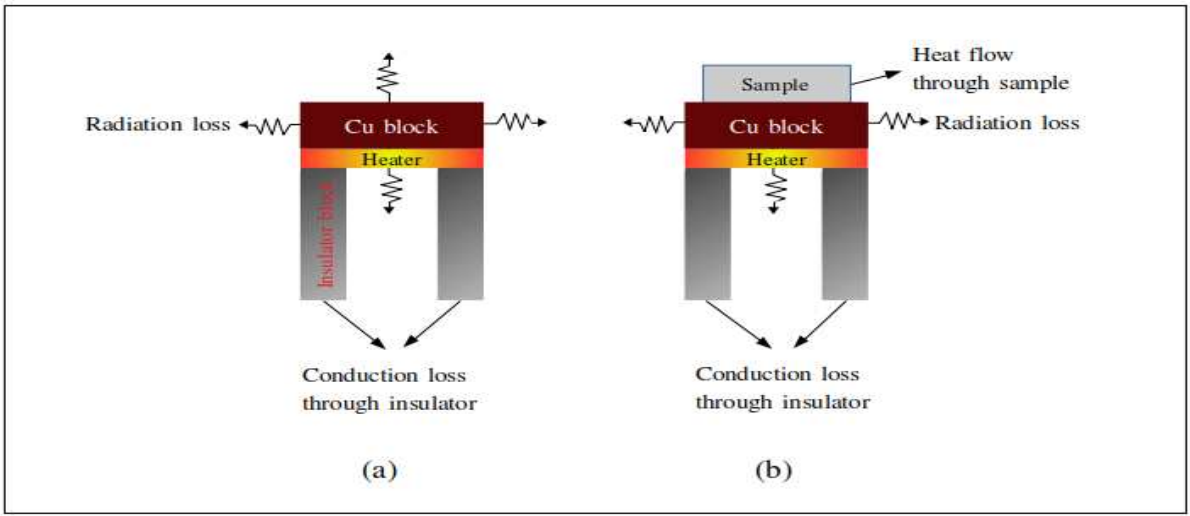} 
\caption{\small{Schematic diagram of heat flow: (a) without sample and (b) with sample.}}
\end{figure*}

\subsection{Measurement of Thermal conductivity}
Steady-state based Fourier's law of thermal conduction is used for the measurement of $\kappa$. In this method, $\kappa$ of a material is calculated as
\begin{equation}
\kappa = \frac{\dot{Q_{s}}}{A}.\frac{l}{\Delta T},
\end{equation}
where $\dot{Q_{s}}$ is the amount of heat passing through the cross-sectional area, $A$ of the sample per unit time. $\Delta T$ is the temperature gradient across the thickness, $l$ of the sample. The accuracy of the measurement of $\kappa$ totally depends on how accurately we measure the $\dot{Q_{s}}$. The difficulty associates with the measurement of $\dot{Q_{s}}$ due to the undefined amount of heat loss during the heat transfer process. In the present work, the amount of heat loss is measured by parallel thermal conductance technique given by Zawilski \textit{et al.}\cite{zawilski} as discussed in Sec. I. To minimize the heat loss we have made a thin heater which is fixed on insulating blocks as shown in Fig. 2.  At the first step, heat loss is recorded by running the instrument without a sample. In this case, heat loss occurs by means of conduction loss through insulator blocks and radiation loss from the surfaces of copper block as shown in Fig. 2(a). For a given input power, the temperature initially increases linearly as a function of time because heat loss is very less. Afterwards, heat loss increases as the temperature increases. Eventually, at equilibrium both the input power and rate of heat loss become equal. At this condition temperature is recorded and the corresponding input power is noted as baseline correction. Then at the next step, sample is placed on a copper block and power delivered to the heater. In this case heat generated by the heater is distributed through insulating blocks as well as through sample as shown in Fig. 2(b). If the same temperature is maintained in both the steps, then input power difference between these two steps (with sample $-$ without sample) will give the net power flows through the sample. Therefore, net heat flow through the sample is mathematically defined as:
\begin{equation}
\dot{Q_{s}} \approx P_{in} - P_{l},
\end{equation} 
where $P_{in}$ is the power delivered to the heater when sample is placed, $P_{l}$ is the baseline power supplied to the heater with empty sample. This $\dot{Q_{s}}$ is used in Eqn. 3 to calculate $\kappa$. At this point, it is important to note that heat loss through side walls of the sample is ignored, since this loss is normally expected to be negligibly small as compared to heat flow through the sample.\cite{differential4}

\section{APPARATUS DESCRIPTION}
A schematic diagram of the instrument is shown in Fig. 3. Different components are labeled by numbers. The sample (5) is placed between the copper blocks (17) \& (18) of cross-sectional area 8 mm $\times$ 8 mm and thickness 3 mm. Two K-type thermocouples (7) \& (8) of 30 swg are welded at the side wall of the copper blocks. Welding is done with the help of fire gun and using thin wire of the special designed material for this purpose having mixture of 95$\%$ silver and 5$\%$ zinc. At the time of welding, if heat is provided to copper block and thermocouple junction for longer time it will make the thermocouple little brittle and may break after few measurements. So, to make it robust we need to cover the thermocouple with glass wool or ceramic tube at welding time. Both the thermocouples are embedded as close as possible to the sample to minimize the unnecessary creation of $\Delta T$ across the copper blocks. Here, it should be mentioned that galinstan (GaInSn) liquid metal eutectic alloy has been used between the copper block and sample to ensure proper thermal contact. Boiling point of galinstan is $\sim$1800 K, which makes it suitable for high temperature application. 

A thin heater (6) of 24 $\Omega$ is used to heat the sample. This heater is made by winding the kanthal wire of 40 swg over a mica sheet of dimension 10 mm $\times$ 8 mm $\times$ 0.2 mm (length $\times$ width $\times$ thickness). This heater is coated with high temperature cement and wrapped by another mica sheet to avoid electrical contact. Then this is covered by copper sheet of thickness 0.4 mm and welded to the copper block (17) over which sample is placed. The copper wire (16) is used to supply the current to the heater. The heater is placed over two insulating gypsum blocks (9) using high temperature cement. The thermal conductivity of gypsum block is very low $\sim$0.017 W/m-K at 300 K and also suitable for high temperature application.\cite{gypsum} The dimension of both the gypsum block is 8 mm $\times$ 2 mm $\times$ 25 mm. A finite gap between two gypsum blocks has been kept to minimize the conduction loss through gypsum block. These gypsum blocks have been made to stand on a brass plate (11). This brass plate is supported with an stainless steel (SS) rod (19) using screw (15). A mica sheet (4) is laid on the cold side copper plate (18) to insulate the sample electrically. An SS rod (2) having threads is used to apply the pressure on the sample, which is screwed with brass plate (3). This brass plate (3) is supported by an SS rod (20) using screw (21). Both the SS rods (19) \& (20) are attached to the SS flange (14). A circular O-ring (10) is used for proper vacuum seal of the chamber. 

The whole assembly of sample holder mounted on the SS flange (14) is covered by an SS made cylindrical vacuum chamber (22) using nut bolts through hole (12). The inner diameter and length of the vacuum chamber are $\sim$10 cm and $\sim$30 cm, respectively. A port (1) of size KF-25 is used to connect the sample chamber to the vacuum pump. The electrical feedthrough (13) is used for electrical communication between inside and outside the vacuum chamber.

\begin{figure}
\includegraphics[width=0.92\linewidth, height=9.0cm]{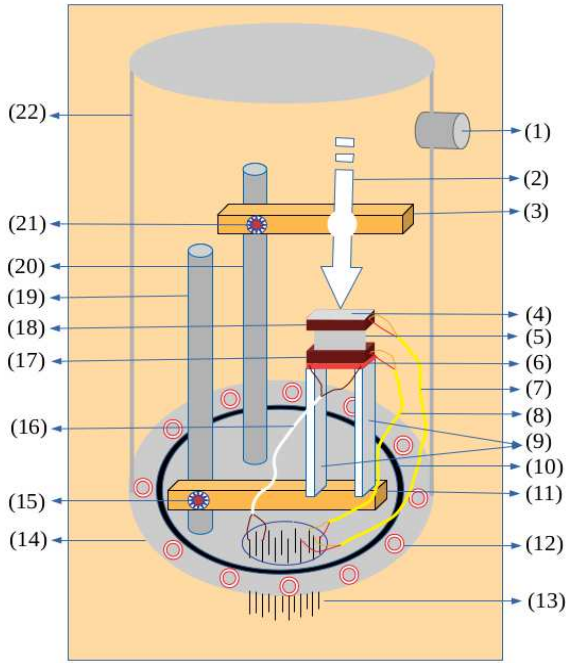} 
\caption{\small{Schematic diagram of the measurement setup.}}
\end{figure}

\section{SAMPLE DIMENSION}
For S measurement sample dimension is not necessary. For $\kappa$, proper sample dimension is required as the thickness and cross-sectional area are used for $\kappa$ measurement. The cylindrical and rectangular shaped samples are used in the present study. Top and bottom surfaces of the sample  should be parallel for the purpose of making good contact of the sample with the copper block. Pellet of the sample should be as compact as possible for the measurement.

\section{MEASUREMENT PROCEDURE}
At first, heat loss measurement has to be done before loading the sample. Once heat loss measurement is performed, the sample (5) is mounted between the copper blocks (17) \& (18). We should ensure that both the surfaces of the sample are in well contact with the copper blocks. A small amount of galinstan are used at the interface of sample and copper block to secure the proper heat transfer. Then vacuum chamber (22) is evacuated using vacuum pump. The Keithley sourcemeter 2604B is used to feed the current to the heater. The various output signals ($T_{1}$, $T_{2}$, $U_{pos}$, $U_{neg}$ etc) are measured by the Keithley digital multimeter 2002 with the multichannel scanner card. The sourcemeter and multimeter are interfaced with computer using IEEE-488B interface bus. It is important to mention that the measurement of S and $\kappa$ for a given sample are carried out simultaneously. 

The whole measurement process is automated with open source programming language "python." Few segments of this program are shown in Fig. 4. PyVISA, NumPy, matplotlib libraries are utilized in the program as displayed in Fig. 4(a). The user only needs to define the input parameters as shown in Fig. 4(b) and then physical interaction of user is not required during the measurement. As per the input power defined, sourcemeter provides current to the heater and wait to achieve steady-state. Steady-state is defined by monitoring the rate of change of $T_{h}$. This rate of change of $T_{h}$ is set as 0.006 K which is checked in every 5 seconds. Program for steady-state is exhibited in Fig. 4(c). Once, the steady-state criteria is fulfilled 5 times successively, multimeter measures the output signals. When the heating cycle reaches its defined maximum temperature limit, heating loop breaks and next loop for cooling starts to cool the sample. During cooling period power in each cycle decreases with defined step size until the sample again comes back to room temperature. Data are also collected in cooling cycle like heating process. Eqns. 2 and 3 are used in the python program to calculate S and $\kappa$, respectively. Here, it is important to note that the temperature dependence of $S_{TC}$ and $S_{neg}$ have been implemented in the program using interpolation method. The raw values of $S_{TC}$ and $S_{neg}$ are taken from the literatures.\cite{boor,bentley} During the measurement, user can observe all the necessary parameters graphically, as the real-time plotting has been employed using matplotlib library. All the measured quantities along with input parameters are saved to [file name].csv file. The live data for acquiring steady-state is shown in Fig. 5. Fig. 5(a) represent the increase in hot side temperature with counts, while Fig. 5(b) exhibits the rate of change of hot side temperature with counts. Every count takes 5 seconds. Hence, for a single data point it takes around 15 minutes.

\begin{figure} 
\begin{subfigure}{0.45\textwidth}
\includegraphics[width=0.6\linewidth, height=2.5cm]{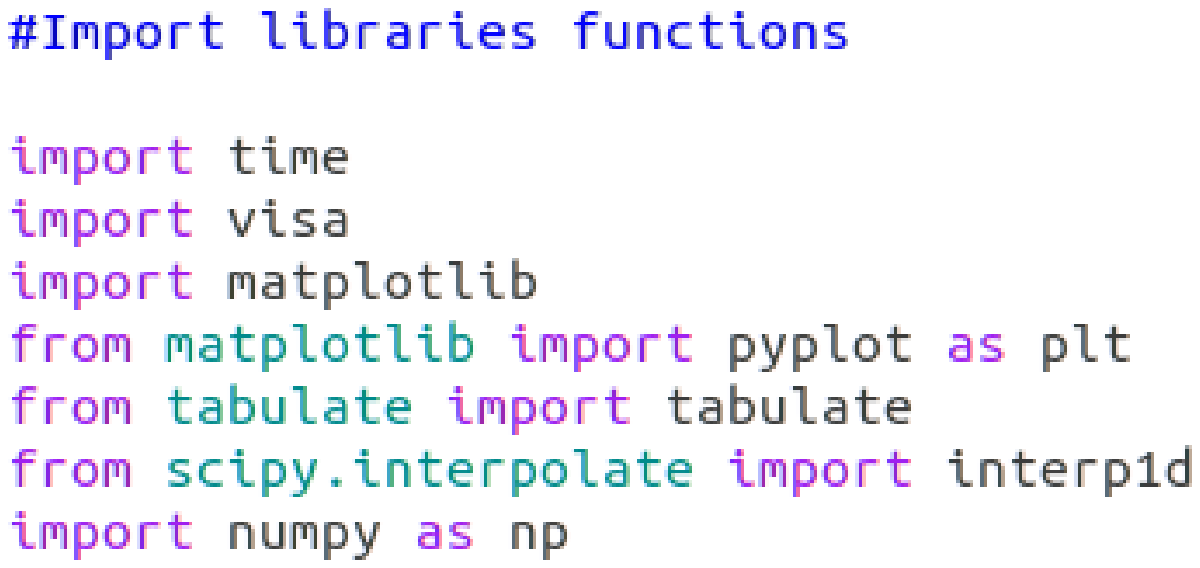} 
\caption{}
\vspace{0.5cm}
\end{subfigure}
\begin{subfigure}{0.45\textwidth}
\includegraphics[width=0.98\linewidth, height=8.8cm]{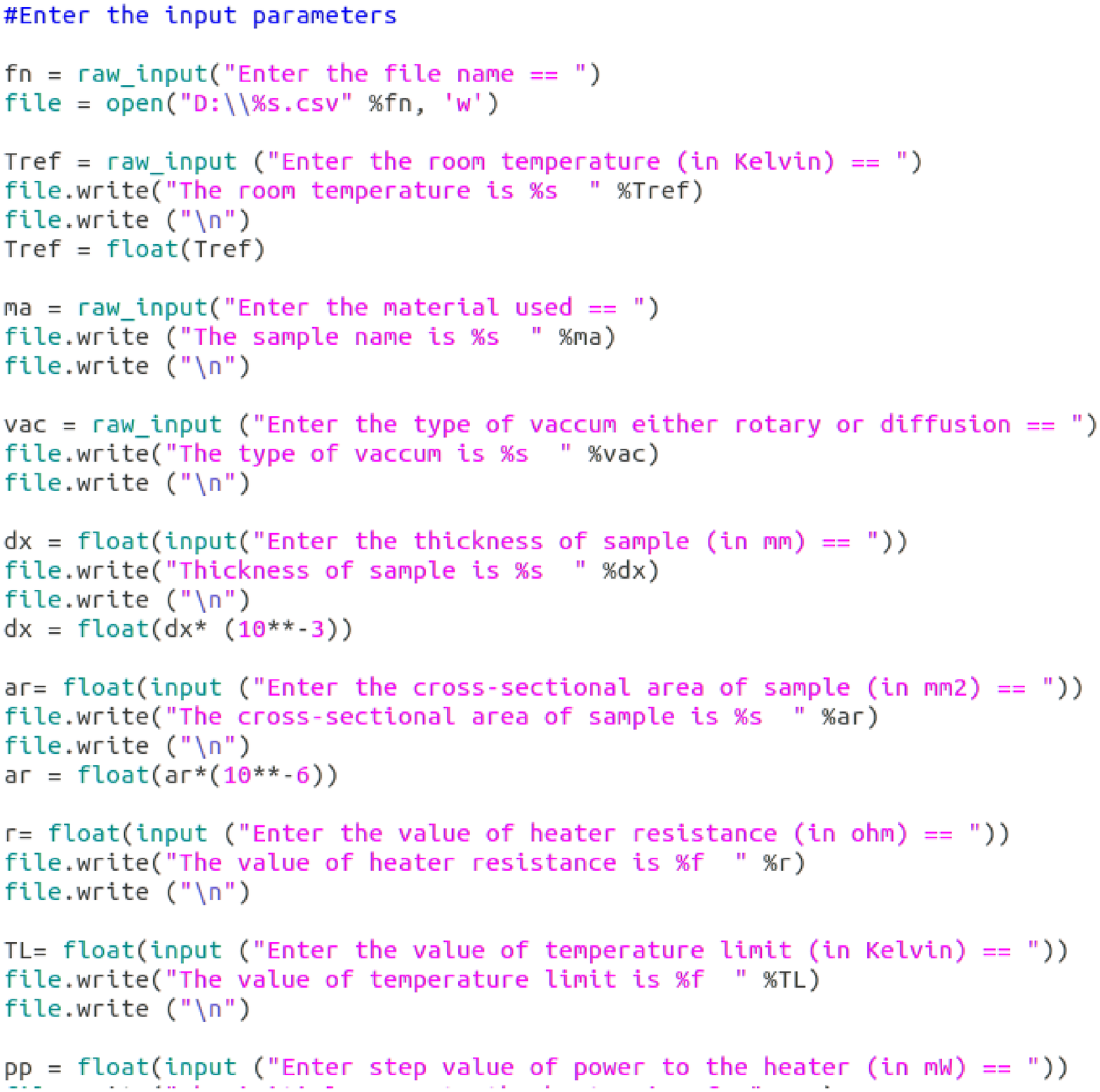}
\caption{}
\vspace{0.5cm}
\end{subfigure} 
\begin{subfigure}{0.45\textwidth}
\includegraphics[width=1.0\linewidth, height=7.3cm]{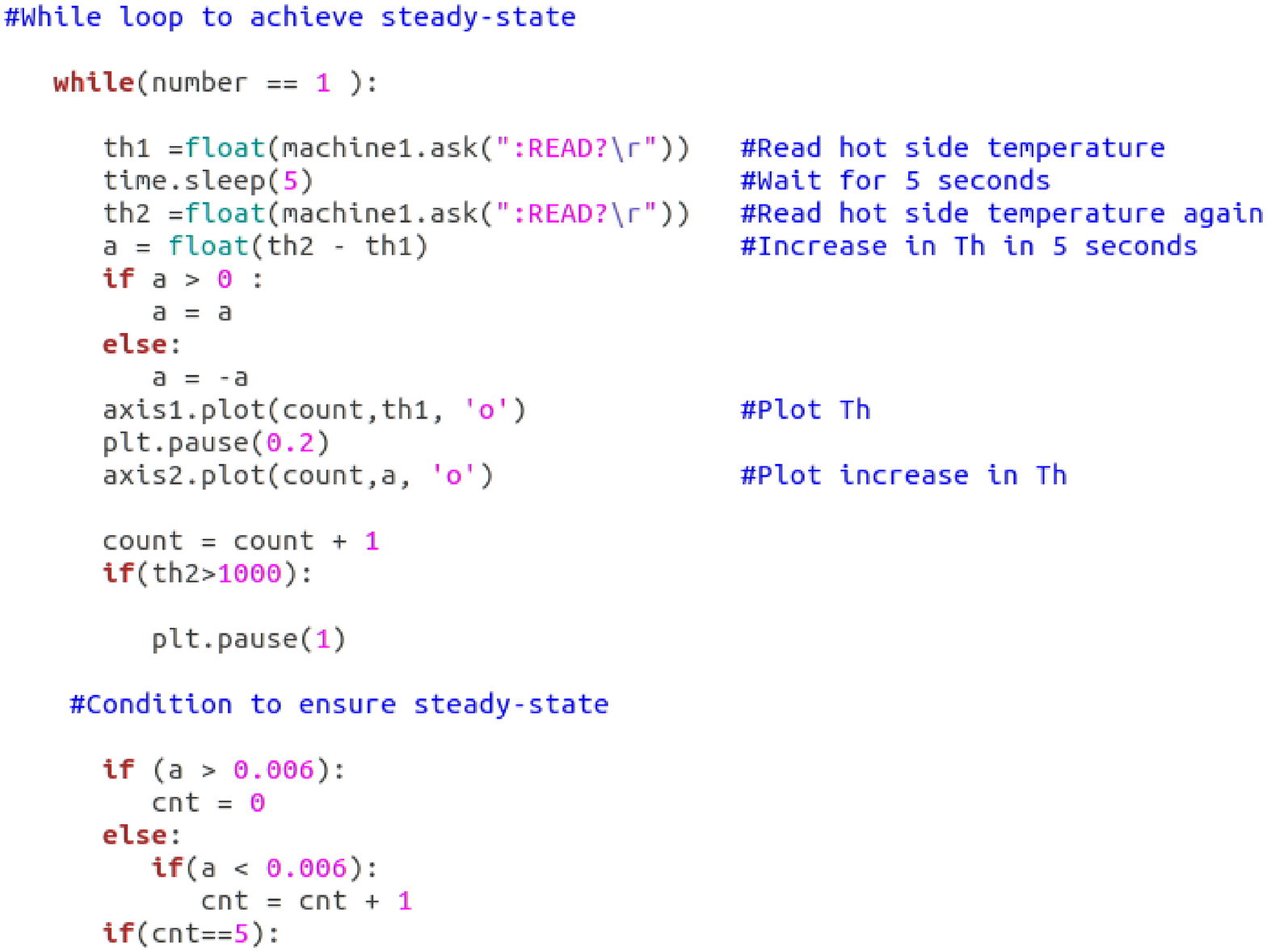}
\caption{}
\vspace{0.3cm}
\end{subfigure} 
\caption{\small{Few segments of python program: (a) Libraries used in python program, (b) input parameters defined by user and (3) program to achieve steady-state.}}
\label{fig:image2}
\end{figure}

\begin{figure*}
\includegraphics[width=0.95\linewidth, height=5.5cm]{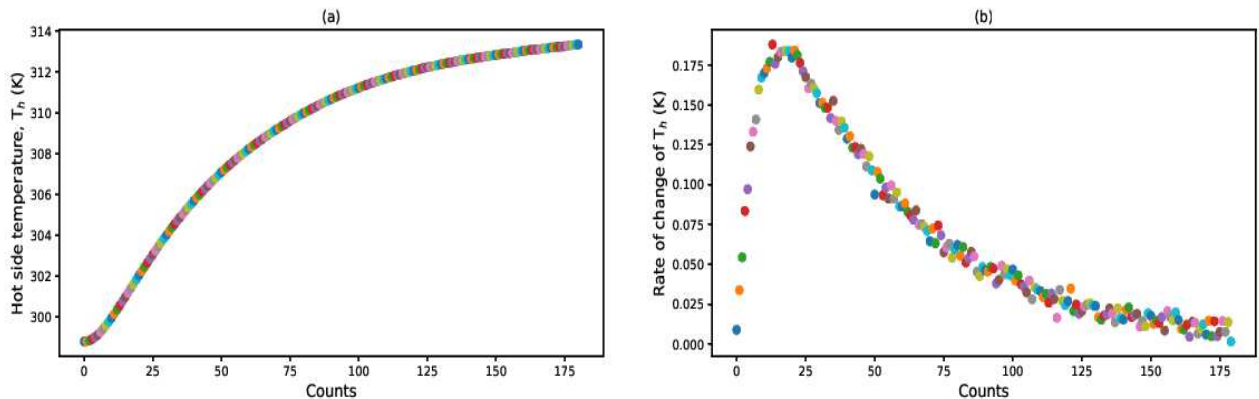} 
\caption{\small{Screenshot of live data: (a) Change in hot side temperature (T$_{h}$) with counts and (b) rate of change of T$_{h}$ with counts. One count takes 5 seconds.}}
\end{figure*}

\section{RESULTS AND DISCUSSION}
\subsection{HEAT LOSS MEASUREMENT}
Before doing the measurement with sample, the heat loss measurement is taken for $\kappa$. Fig. 6 shows the heat loss as a function of hot side temperature ($T_{h}$). The reliability of $\kappa$ value is desperately depend on how accurately measure the heat loss value. Keeping this in mind, we have made a thin heater which has been placed on insulating gypsum blocks to minimize the heat loss and to achieve required temperature easily. Fig. 6 shows heat loss is quite low at the beginning. As the temperature increases heat loss also increases non-linearly. This non-linear increment is fitted with polynomial of quartic degree: Heat loss = $aT + bT^{4}$. Where, values of $a$ and $b$ are 0.959 $mW/K$ and 1.566 $\times$ 10$^{-8}$ $mW/K^{4}$, respectively. The first term indicates the convection loss where heat loss depends on temperature linearly. The quartic term signifies the radiation loss at high temperature region which confirms by Stefan's law: $E = \sigma T^{4}$ (E is radiant heat energy and $\sigma$ is Stefan's constant). The purpose of this polynomial fit is that (i) user can get heat loss at any value of temperature in between $300-800$ K, (ii) the heat loss value can also be calculated at little above 800 K.  

\begin{figure}
\includegraphics[width=0.95\linewidth, height=7.0cm]{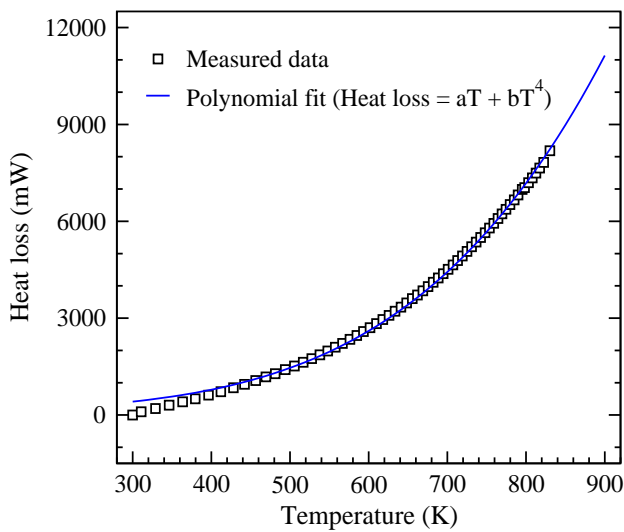} 
\caption{\small{Heat loss at different value of hot side temperature.}}
\end{figure} 
   
\subsection{SEEBECK COEFFICIENT MEASUREMENT}
The instrument is validated by measuring S of different samples in the temperature region $300-800$ K. In order to check the flexibility of the instrument, measurement of various class of samples having wide range of S are carried out. Fig. 7 shows the measured S of nickel, Fe$_{2}$VAl and LaCoO$_{3}$ as a function of temperature. The shape and dimension of these samples are tabulated in Table 1. 

Fig. 7(a) shows the measured S as a function of temperature of high purity (99.99$\%$) nickel. The value of S at 300 K is found to be $\sim-$19.2 $\mu$V/K, then decreases up to $\sim$490 K with the corresponding value of $\sim-$24.4 $\mu$V/K. After 490 K, the S increases monotonically till $\sim$660 K and has the value of $\sim-$18.9 $\mu$V/K. The value of S continuously decreases above 660 K and reaches $\sim-$22.9 $\mu$V/K at 800 K. In order to verify, we have compared our S with the reported data of Burkov \textit{et al.}\cite{burkov} and Ponnambalam \textit{et al.}\cite{ponnambalam} in the same figure. The S value matches well with the reported data with the maximum deviation of $\approx$ 1 $\mu$V/K in the whole studied temperature range.

\begin{table}
\caption{\small{Shape and dimension of the test samples.}}
\resizebox{0.48\textwidth}{!}{%
\begin{tabular}{@{\extracolsep{\fill}}l c c c c c c c c c c c c c c c} 
\hline\hline
 
\multicolumn{1}{c}{Sample name} &&& \multicolumn{1}{c}{Shape of} &&& \multicolumn{4}{c}{Thickness} &&& \multicolumn{4}{c}{Cross-sectional area} \\                                 
  &&& \multicolumn{1}{c}{cross-section} &&&&& \multicolumn{1}{c}{(mm)}   &&&&&&&   \multicolumn{1}{c}{(mm$^{2}$)} \\
   
\hline
Nickel         &&& Circular    &&&&& 2    &&&&&&& 28.27  \\
Gadolinium     &&& Rectangular &&&&& 0.7  &&&&&&& 4.86  \\
Fe$_{2}$VAl    &&& Rectangular &&&&& 1.38 &&&&&&& 15  \\
LaCoO$_{3}$    &&& Rectangular &&&&& 0.5  &&&&&&& 24  \\ [0.5ex]
 
\hline\hline
 
\end{tabular}}
\end{table} 

\begin{figure*}
\includegraphics[width=1.0\linewidth, height=5.4cm]{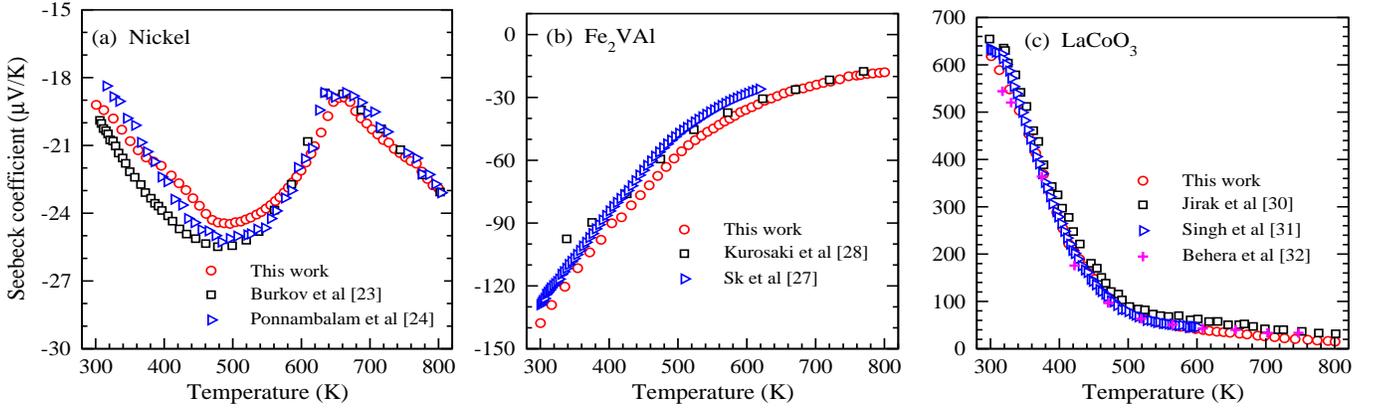} 
\caption{\small{Seebeck coefficient of (a) Nickel, (b)Fe$_{2}$VAl and (c) LaCoO$_{3}$ as a function of temperature.}}
\end{figure*}

Fig. 7(b) exhibits the temperature dependent S of Fe$_{2}$VAl. This compound possesses a deep pseudogap at the Fermi level and due to this it is known to give the large S and $\sigma$.\cite{nishino, hanada} This compound has been synthesized by arc melting technique.\cite{shamim_mrx} The value of S is observed as $\sim-$138 $\mu$V/K at 300 K. The S increases monotonically up to the highest temperature of interest. At 800 K, S is found to be $\sim-$18 $\mu$V/K. Our result is compared with the reported data by Sk \textit{et al.}\cite{shamim_mrx} and Kurosaki \textit{et al.}\cite{kurosaki} as shown in the same figure. One can notice that our data has good agreement with the reported data throughout the temperature range. However, S of Kurosaki \textit{et al.}\cite{kurosaki} and our data get deviating at low temperature region where this deviation vanishes as the temperature increases. This deviation may be due to the different annealing temperature were applied in synthesizing the compound in the respective works.

The S of LaCoO$_{3}$ is measured in the temperature region $300-800$ K as displayed in Fig. 7(c). This compound is synthesized by combustion technique. LaCoO$_{3}$ gives a large S at 300 K\cite{maignan, jirak}, which is a good signature for efficient TE material. At 300 K, S is found to be $\sim$618 $\mu$V/K then decreases continuously throughout the full temperature range as Fig 7(c) shows. At 800 K, the observed value of S is $\sim$15 $\mu$V/K. Figure shows that S is decreasing rapidly from 300 to $\sim$500 K. After 500 K, this decrement rate falls down and continues up to 800 K. In the temperature interval $\sim300-500$ K, the change in S is $\sim$535 $\mu$V/K with decrement rate of 2.67 $\mu$V/K. From 500 to 800 K, this change in S is observed as $\sim$68 $\mu$V/K with the decrement rate of 0.23 $\mu$V/K. The similar trend of S is reported by other groups\cite{jirak, singh, behera} also, which are shown in the same Fig. 7(c). In the full studied temperature range, our S value give the good match with reported data.

\begin{figure*}
\includegraphics[width=1.0\linewidth, height=5.4cm]{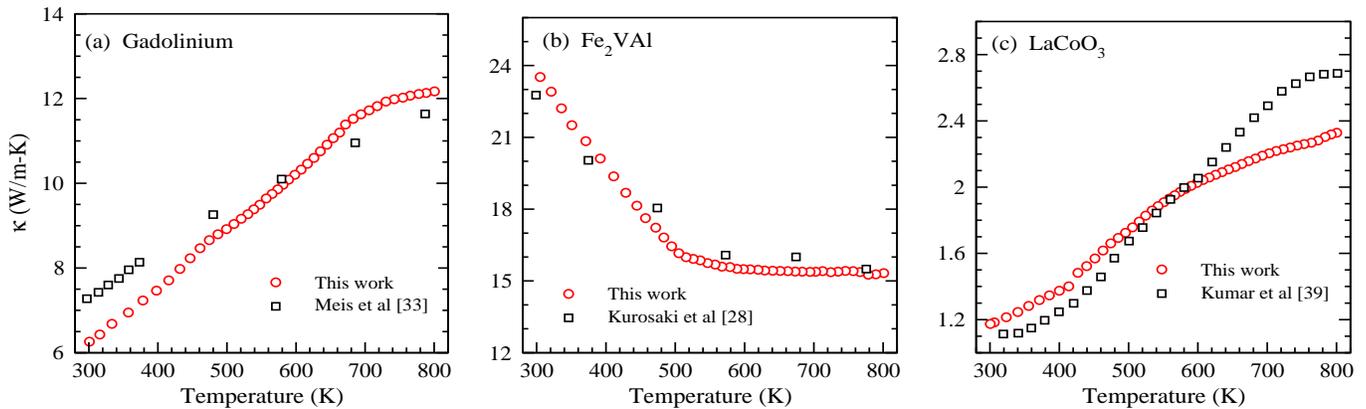} 
\caption{\small{Temperature dependence of thermal conductivity of (a) Gadolinium, (b)Fe$_{2}$VAl and (c) LaCoO$_{3}$.}}
\end{figure*} 

Among the test samples used here, nickel is a pure metal, Fe$_{2}$VAl is an inter-metallic Heusler-type compound and LaCoO$_{3}$ is an oxide perovskite material. Hence, measurement of S on various kind of materials are carried out in the present study. This suggests that our setup can be used to measure S in the temperature range $300-800$ K for any class of materials having wide range of S values.  

\subsection{THERMAL CONDUCTIVITY MEASUREMENT}
To examine our setup, we have measured $\kappa$ of gadolinium, Fe$_{2}$VAl and LaCoO$_{3}$ in the temperature range $300-800$ K as shown in Fig. 8. The shape and dimension of these samples are shown in Table 1. Thickness and cross-sectional area of the sample are important for the measurement of $\kappa$.

Fig. 8(a) represents the measured $\kappa$ for gadolinium as a function of temperature. Gadolinium has one of the lowest $\kappa$ among the metallic compounds.\cite{meis} From Fig. 8(a), the $\kappa$ is observed to increase with increasing temperature. The values are found to be $\sim$6.3 and $\sim$12.2 W/m-K at 300 and 800 K, respectively. Our $\kappa$ is compared with the reported data of Meis \textit{et al.}\cite{meis} as shown in the same figure. The $\kappa$ of gadolinium at room temperature is highly controversial. Because around room temperature gadolinium possesses the critical ferromagnetic (16$^{o}$ C) and paramagnetic ($25-29^{o}$ C) phases.\cite{pascal} In $18-25^{o}$ C, the values of $\kappa$ measured using different methods are reported from 6.6 to 14 W/m-K in the different literaure.\cite{meis,legvold,arajs,powell} Fig. 8(a) shows that at lower temperature region both the data are little deviating, while this deviation diminishes at high temperature region.     

Fig. 8(b) illustrates the temperature dependence values of $\kappa$ for Fe$_{2}$VAl. The value of $\kappa$ at 300 K is found to be $\sim$ 23.5 W/m-K. From temperature 300 to $\sim$500 K $\kappa$ decreases drastically then $\kappa$ is almost constant throughout the remaining temperature range. At 800 K, the observed value of $\kappa$ is $\sim$ 15.3 W/m-K. In the temperature range $300-500$ K, the magnitude change in $\kappa$ is 7 W/m-K, whereas this value is 1 W/m-K in the temperature range $500-800$ K. Our $\kappa$ is well matched with the reported value of Kurosaki et al.\cite{kurosaki} in the whole temperature region as shown in the figure. 

We have also carried out the measurement of $\kappa$ for the oxide sample. Oxide sample is known for having low $\kappa$. The temperature dependence of $\kappa$ for LaCoO$_{3}$ is shown in Fig. 8(c). The value of $\kappa$ is measured as $\sim$1.2 W/m-K at 300 K, then increases monotonically up to 800 K with the corresponding value of $\kappa$ is $\sim$2.3 W/m-K. Figure shows that the $\kappa$ increases almost linear up to $\sim$500 K then get saturated as the temperature increases. The values of $\kappa$ around room temperature are reported from 0.38 to 3.2 W/m-K in the different works.\cite{cortes,behera,kumar,li,bousnina,pillai} As we know that $\kappa$ depends on particle size as well as porosity of the sample. Therefore, this diference is may be because of sample has been synthesized by different techniques and $\kappa$ is measured by various methods in the respective works.\cite{cortes,behera,kumar,li,bousnina,pillai} Our $\kappa$ is closely matched with the reported value of Kumar \textit{et al.}\cite{kumar} as shown in the figure. A small deviation is observed at high temperature. Maximum deviation of 0.36 W/m-K is found at 800 K.

The various type of samples are tested for $\kappa$ measurement using our instrument with wide range of $\kappa$ from $\sim$1.1 to $\sim$23.5 W/m-K. Our instrument is also examined for wide range of S from $\sim-$20 to $\sim$600 $\mu$V/K. Therefore, this study propose that present apparatus can be used for the simultaneous measurement of S and $\kappa$ in the temperature range $300-800$ K.

\section{CONCLUSIONS}
In conclusion, we have fabricated the instrument for measuring Seebeck coefficient and thermal conductivity simultaneously in the temperature range $300-800$ K. A simple, low-cost and user friendly instrument is fully automated with open source programming language python. The differential method is used for Seebeck coefficient measurement, where the same thermocouple is employed to measure temperature as well as voltage. Temperature dependent Seebeck coefficient of thermocouple has been taken care. The parallel thermal conductance technique is implemented for thermal conductivity measurement. We have made a thin heater and insulating heater base to minimize the heat loss. The instrument is validated by doing measurement on various kind of samples with different dimensions and shapes. The measurement on test samples like nickel, gadolinium, Fe$_{2}$VAl and LaCoO$_{3}$ are carried out. The instrument is tested for a wide range of Seebeck coefficient from $\sim-$20 to $\sim$600 $\mu$V/K and thermal conductivity from $\sim$1.1 to $\sim$23.5 W/m-K. The measured values are found in good agreement with the reported values.
   
\vspace{0.5cm}
\textbf{Data Availability Statement}

Data available on request from the authors.

\end{document}